\documentclass[12pt]{iopart}

\usepackage{amssymb}
\usepackage[dvips]{graphicx}

\newcommand{\ep}{\varepsilon}

\newcommand{\sit}{\sin^2{\theta}}
\newcommand{\sid}{\sin^3{\theta}}

\newcommand{\sinul}{\sin{\theta_0}}
\newcommand{\sitnul}{\sin^2{\theta_0}}

\newcommand{\sitgem}{\langle\sin^2{\theta}\rangle}
\newcommand{\sidgem}{\langle\sin^3{\theta}\rangle}

\newcommand{\sitgeme}{\langle\sin^2{\theta}\rangle_{_1}}
\newcommand{\sidgeme}{\langle\sin^3{\theta}\rangle_{_1}}
\newcommand{\sitgemt}{\langle\sin^2{\theta}\rangle_{_2}}
\newcommand{\sidgemt}{\langle\sin^3{\theta}\rangle_{_2}}
\newcommand{\singeme}{\langle\sin^n{\theta}\rangle_{_1}}
\newcommand{\singemt}{\langle\sin^n{\theta}\rangle_{_2}}

\newcommand{\su}{\sum_i \rho_i S_i}

\newcommand{\Te}{T_{\mathrm{eff}}}

\newcommand{\sgn}{\mathrm{sgn}}

\newcommand{\Vgem}{\langle V\rangle}

\begin{document}

\title{Thermal Brownian motor}

\author{P Meurs and C Van den Broeck}

\address{Hasselt University, B-3590 Diepenbeek, Belgium}

\begin{abstract}
Recently, a thermal Brownian motor was introduced [Van den Broeck,
Kawai and Meurs, {\it Phys. Rev. Lett.}, 2004], for which an exact
microscopic analysis is possible. The purpose of this paper is to
review some further properties of this construction, and to
discuss in particular specific issues including the relation with
macroscopic response and the efficiency at maximum power.
\end{abstract}


\submitto{\JPCM}

\section{Introduction}

Brownian motors are small scale engines that operate on the basis
of nonequilibrium fluctuations. Most of the models that have been
discussed previously in the literature refer to the transfer of
chemical or potential energy into work \cite{reimann} by Brownian
entities that operate in a externally imposed asymmetric
environment (or by  spontaneous symmetry breaking \cite{prost}).
In this case, a $100$ percent  efficiency can in principle be
achieved \cite{parrondo}, in accordance with the basic laws of
thermodynamics. Recently, a thermal Brownian motor was introduced
\cite{vandenbroeckprl}, for which an exact microscopic analysis
turns out to be possible \cite{meurs}. The model can be viewed as
a simplification of the Feynman-Smoluchowski ratchet
\cite{smoluchowski,feynman}. This construction differs in two
essential aspects from most other Brownian motors. First, the
source of energy is the presence of different heat baths.
Thermodynamics thus predicts that the limiting efficiency is that
of the Carnot engine. Second, the asymmetry is an intrinsic
property of the motor, not of the environment or of the imposed
constraint. The purpose of this paper is to review some further
properties of this construction, and to discuss in particular
specific issues including the relation with macroscopic response
and the efficiency at maximum power.

\section{Model and basic results}

\begin{figure}
\begin{center}
\includegraphics[width=12cm]{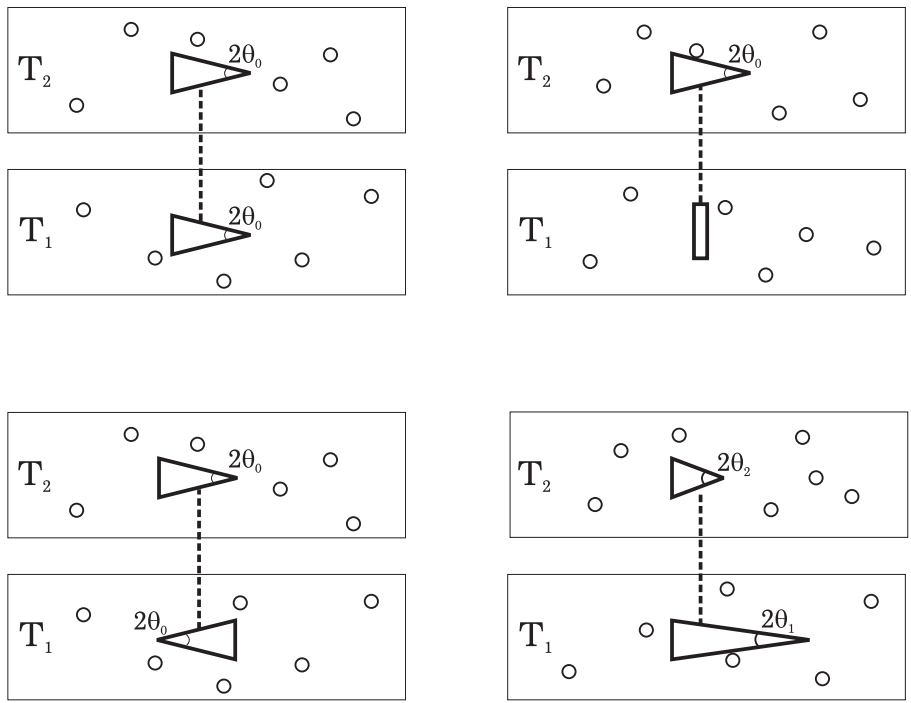}
\caption{Four different realizations of the thermal Brownian
motor, each consisting of two rigidly linked motor units. (a) will
be referred to as Triangula and is composed out of two identical
triangles, while model (b) is called Triangulita and has a
triangle and a bar. Constructions (c) and (d) consist respectively
out of two identical triangles pointing in the opposite direction
and two triangles with a different apex angle.}\label{fig:models}
\end{center}
\end{figure}

\begin{figure}
\begin{center}
\includegraphics[width=5cm]{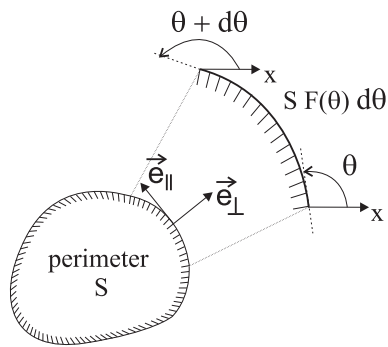}
\end{center}
\caption{A closed and convex object with perimeter $S$. The length
of the surface with an orientation between $\theta$ and
$\theta+d\theta$ is $SF(\theta)d\theta$, defining the form factor
$F(\theta)$.}\label{fig:shape}
\end{figure}

We first briefly review the basic construction of the thermal
Brownian motor and the main results. The motor consists of a set
of convex units $i$ ($i=1,...,N$), that move as a rigidly linked
entity with a single degree of freedom, along a specified
direction $x$. Each unit  resides in a separate, infinitely large
compartment that is filled with an ideal gas at equilibrium. These
compartments play a role akin to the thermal reservoirs in a
Carnot construction. For simplicity, we consider the case of
spatial dimension $d=2$ with coordinates $(x,y)$. In Fig.
\ref{fig:models} we show four different realizations of the motor.
The gas particles are point particles of mass $m$. Their density,
temperature and Maxwellian velocity distribution in compartment
$i$ are denoted by $\rho_i$, $T_i$ and $\phi_i$ respectively,
with:
\begin{eqnarray}
\phi_i(v_x,v_y)=\frac{m}{2 \pi k_B
T_i}\exp{\left(\frac{-m(v_x^2+v_y^2)}{2 k_B T_i}\right)}.
\end{eqnarray}
We consider the limit in which the mean free path of the gas
particles is much larger than the linear dimension of the motor
unit. In this case, there are no pre-collisional correlations
between the speed of the motor and that of the impinging
particles. The probability density $P(V,t)$  for the speed
$\vec{V}=(V,0)$ therefore obeys a Boltzmann-Master equation:
\begin{eqnarray}\label{boltzmann}
\frac{\partial P(V,t)}{\partial t}=\int{dV'
\Big[W(V|V')P(V',t)-W(V'|V)P(V,t)\Big].}
\end{eqnarray}
$W(V|V')=\sum_i W_i(V|V')$ is the transition probability per unit
time for the motor to change speed from $V'$ to $V$ due to the
collisions with the gas particles in the compartments
$i=1,\ldots,N$. Its explicit form, following from the collision
rules and basic arguments from kinetic theory, is given by (see
\cite{meurs} for more details):
\begin{eqnarray}
\fl W_i(V|V') = \int_0^{2\pi} d\theta\, S_i F_i(\theta)
  \int_{-\infty}^{+\infty} dv_x^\prime
  \int_{-\infty}^{+\infty} dv_y^\prime
\rho_i \phi_i(v_x^\prime,v_y^\prime) ( \vec{V}'-\vec{v}' ) \cdot
\hat{e}_\perp
\nonumber \\
\times \Theta\left [( \vec{V}'-\vec{v}' ) \cdot \hat{e}_\perp
\right ]
 \delta \left [
   V - V^\prime -
   \frac{2 \frac{m}{M} \sin^2 \theta}{1+\frac{m}{M} \sin^2 \theta}
   \left ( v_x^\prime - v_y^\prime \cot\theta -V^\prime  \right )
   \right ].
\label{eq:W}
\end{eqnarray}
Here $\Theta$ is the Heaviside step function. The $\delta$ Dirac
function selects appropriate particle velocities yielding the
post-collisional speed $V$ for the motor in function of the
precollisional speeds $V'$ and $v'_x,v'_y$ of motor and gas
particle respectively. The shape of the motor unit $i$ is
characterized by a normalized shape function $F_i(\theta)$, such
that $F_i(\theta) d\theta$ is the fraction of the outer surface of
the unit  that has a tangent between $\theta$ and $\theta +
d\theta$, cf. Fig.~\ref{fig:shape}. The angle $\theta$ is measured
counterclockwise from the horizontal $x$ direction. The total
perimeter will be denoted by $S_i$.

Solving the Boltzmann-Master equation in general is out of the
question, even at the steady state. To make progress, a
perturbative solution is necessary. Since we expect that the
rectification disappears in the limit of a macroscopic motor, a
natural expansion parameter is the ratio of the mass $m$ of the
gas particle over the mass $M$ of the object. More precisely, we
have used $\ep=\sqrt{m/M}$ as the expansion parameter.

Introducing the transition probability $W_i(V';r)=W_i(V|V')$ with
jump amplitude $r=V-V'$, the Boltzmann equation can be rewritten
as
\begin{eqnarray}\label{masterjump}
\frac{\partial P(V,t)}{\partial
t}&=&\int{W(V-r;r)P(V-r,t)dr}-P(V,t)\int{W(V;-r)dr}.
\end{eqnarray}
The Taylor expansion of the r.h.s. of this equation with respect
to the jump amplitude leads to the equivalent Kramers-Moyal
expansion:
\begin{eqnarray}
\frac{\partial P(V,t)}{\partial
t}=\sum_{n=1}^{\infty}{\frac{(-1)^n}{n!}
\left(\frac{d}{dV}\right)^n\left\{a_n(V)P(V,t)\right\},}\label{kramers-moyal}
\end{eqnarray}
with  jump moments $a_n(V)=\sum_i a_{n,i}(V)$, defined as:
\begin{eqnarray}\label{defAn}
a_{n,i}(V)=\int{r^n W_i(V;r)dr}.
\end{eqnarray}

One can obtain  from the Kramers-Moyal expansion the  set of
coupled equations for the moments of the velocity distribution
$\langle V^n\rangle=\int V^n P(V,t)$. Of interest for the
subsequent discussion, we reproduce the exact results for the
first two moment equations:
\begin{eqnarray}
\partial_t\langle V\rangle &=&\langle a_1(V)\rangle \label{seteq}\\
\partial_t \langle V^2\rangle&=&2\langle V a_1(V)\rangle + \langle
a_2(V)\rangle \label{seteq2},
\end{eqnarray}
with
\begin{eqnarray}
\fl a_{1,i}(V)=-\rho_i S_i
\int_0^{2\pi}d\theta F_i(\theta)\frac{1}{M+m\sin^2{\theta}}\nonumber\\
 \left[\sqrt{\frac{2 k_B T_i m}{\pi}}\exp[-\frac{m}{2 k_B
T_i}V^2\sin^2{\theta}]V \sin^2{\theta}+\right.\nonumber\\
 +\left. \left(1+\textrm{Erf}\left[\sqrt{\frac{m}{2 k_B
T_i}}V\sin{\theta}\right]\right)\left(k_B T_i \sin{\theta} + m V^2
\sin^3{\theta} \right)\right]\label{a1}
\end{eqnarray}
\begin{eqnarray}
\fl a_{2,i}(V)=2\rho_i S_i
\int_0^{2\pi}d\theta F_i(\theta)\frac{\sin^2\theta}{(M+m\sin^2{\theta})^2}\nonumber\\
 \left[\sqrt{\frac{2 k_B T_i m}{\pi}}\exp[-\frac{m}{2 k_B
T_i}V^2\sin^2{\theta}]\left(2k_B T_i+m V^2 \sin^2{\theta}\right)
+\right.\nonumber\\
+\left. \left(1+\textrm{Erf}\left[\sqrt{\frac{m}{2 k_B
T_i}}V\sin{\theta}\right]\right)\left(3 k_B T_i m V \sin{\theta} +
m^{2} V^3 \sin^3{\theta} \right)\right].\label{a2}
\end{eqnarray}

\section{Systematic speed: general results}

\begin{figure}
\begin{center}
\includegraphics[width=10cm]{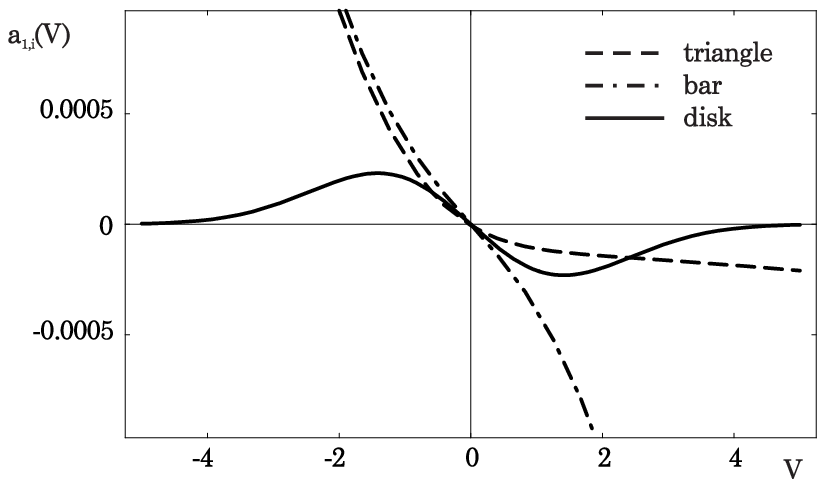}\vspace{0.5cm}
\includegraphics[width=9.7cm]{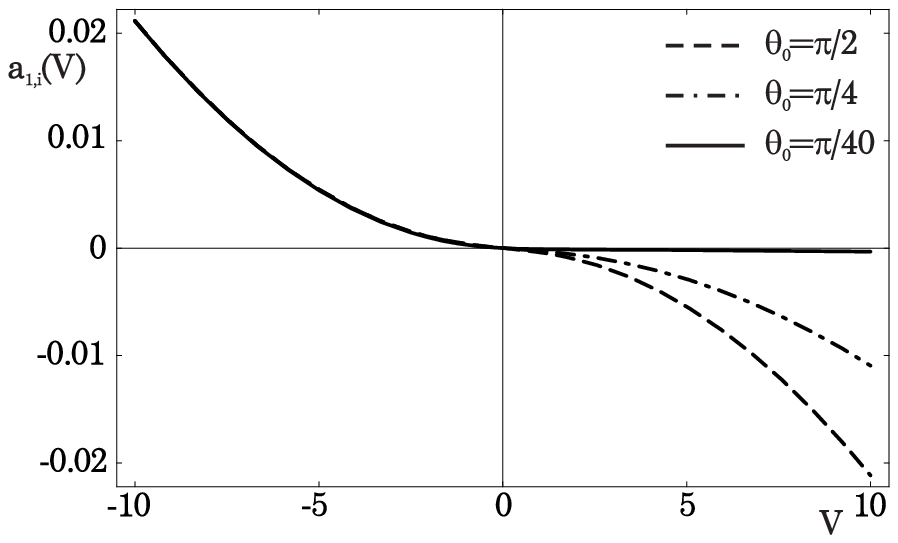}
\caption{({\bf a}) {\it Top}: The response function $a_{1,i}(V)$,
see Eq. (\ref{a1}), for different geometric objects used in de
constructions of Fig. \ref{fig:models}: a triangle with
$\theta_0=\pi/36$, a bar of length $L$ and a disk with radius $R$.
 \\
({\bf b}) {\it Bottom}: The function $a_{1,i}(V)$ for a triangle
with opening angle respectively equal to $\pi, \pi/2$ and
$\pi/20$. The following parameter values were used: mass ratio
$M/m=20$, density $\rho_i=0.0022$, temperature $T_i=1$, size
$L=R=1$ and $k_B=1$ by choice of units.}\label{fig:a1}
\end{center}
\end{figure}

The equation for the first moment is coupled to the higher
moments, a familiar feature in statistical mechanics. However, a
perturbative solution can be obtained by considering the Taylor
expansion in $\ep$. To lowest order in $\ep$ the first moment
equation reduces to a closed equation featuring a linear
relaxation law, namely:
\begin{eqnarray}
M\partial_t\Vgem=-\gamma \langle V\rangle,
\end{eqnarray}
where
$\gamma=\sum_i\gamma_i$ and $\gamma_i$ the linear friction
coefficient, due to the section of the motor sitting in gas
mixture $i$:
\begin{eqnarray}
\gamma_i=4\rho_i S_i \sqrt{\frac{k_BT_im}{2\pi}}\int_0^{2\pi}
{d\theta F_i(\theta)\sit}.\label{gammageneral}
\end{eqnarray}
Note that at this level of the perturbation, the steady state speed of the motor is
zero: even though the motor may be constituted of asymmetric parts, no rectification takes place at the level of linear
response.

At the next order in the perturbation, the first moment is coupled to the second one,
 which needs to be solved at lowest order.
By doing so, one finds at the steady state:
\begin{eqnarray}
M\langle V^2\rangle=k_B \Te ,
\end{eqnarray}
 with  $\Te$ the effective temperature given by:
\begin{eqnarray}
\Te=\frac{\sum_i{\gamma_i T_i}}{\sum_i{\gamma_i}}.\label{teff}
\end{eqnarray}
By inserting this result in the equation for the average velocity,
one finds for the lowest order term of the steady state velocity
$V_0$:
\begin{eqnarray}
V_0\equiv\langle V\rangle= \sqrt{\frac{m}{M}}\sqrt{\frac{\pi
k_B\Te}{8M}} \frac{\su \left(\frac{T_i}{\Te} - 1\right)
\int_0^{2\pi}d\theta F_i (\theta) \sin^3 \theta}{\su
\sqrt{\frac{T_i}{\Te}} \int_0^{2\pi} d\theta F_i(\theta)
\sin^2(\theta)}.\label{eq:general}
\end{eqnarray}
This speed is equal to the expansion parameter times the thermal
speed of the motor and further multiplied by a factor that depends
on the geometric properties of the object. Note that the Brownian
motor ceases to function in the absence of a temperature
difference (when $T_i = T_{\mathrm{eff}}$ for all $i$) and in the
macroscopic limit $M\rightarrow \infty$ (since $V_0 \sim 1/M$).
Note also that the speed is scale-independent, i.e., independent
of the actual size of the motor units: $V_0$ is invariant under
the rescaling $S_i$ to $C S_i$. In Table \ref{tab:models} we have
collected the explicit result for $V_0$ for the four different
thermal motors, depicted in Fig. \ref{fig:models}.

It is tempting to interpret the direction of the systematic speed
on the basis of intuition. One can study the response of the motor
by the application of an external force. Due to the asymmetry, the
response curve (speed versus force) will be asymmetric. In
particular, it will require less force to move a triangle in the
``easy'' direction of the sharp angle. Consequently, one expects
that fluctuations leading to an ``extra force''  in the easy
direction would be more successful in moving the object than the
ones in the other direction. A similar argument can be applied to
the discussion of the electric current in a diode, which also has
an asymmetric response of current versus field. One expects that
fluctuations in the electric field will produce a larger current
in one direction than in the other one. However, as is well known
\cite{vankampen}, these arguments are clearly wrong. The first
observation is that there is no systematic speed (or electric
current) when the system is at equilibrium. In fact, the effect of
possible asymmetries in a system are completely wiped out in
equilibrium, as expressed explicitly in the principle of detailed
balance discovered by Onsager \cite{onsager}: at equilibrium any
process and its time-reverse occur equally frequently. Second,
equilibrium is  typically a so-called point of flux reversal
\cite{feynman}. For example, in the case of Triangulita, the flux
is in the ``easy'' direction of the sharp angle, only in the case
when the temperature is lower in the reservoir containing the
triangle. Otherwise the speed is in the opposite direction.
Furthermore the speed again becomes zero when the temperature is
zero in the reservoir containing the flat paddle. Finally, the
above mentioned asymmetry in the response appears in the nonlinear
regime. In this regime, the dynamics of the various moments are
however coupled. In particular, the equation for the average speed
is not closed, but depends on the entire probability distribution.
The nonlinear response to an external force is therefore not
linked in an obvious way to the intrinsic nonlinear dynamics
generated by nonequilibrium fluctuations.

Full information on the equation of the first moment is contained
in the nonlinear function  $a_{1,i}(V)$. It is reproduced in
figures \ref{fig:a1}(a) and \ref{fig:a1}(b) for the motor units
used in the constructions of Fig. \ref{fig:models}. However, the
direction of motion, to lowest order in $\ep$, contains much less
information. Indeed, from Eq. (\ref{eq:general}), one finds for a
motor with $2$ units:
\begin{eqnarray}\label{sgn}
\sgn(V_0)=\sgn\left[(T_1-T_2)\left(\frac{\sidgeme}{\sitgeme}
-\sqrt{\frac{T_1}{T_2}}\frac{\sidgemt}{\sitgemt}\right)\right].
\end{eqnarray}
Hence the direction of the average speed does not depend on the
densities of the gases or the size of the motor units. The
quantity
$-1\leq\langle\sin^3\theta\rangle_{_i}=\int_0^{2\pi}d\theta
F_i(\theta)\sin^3\theta\leq 1$ is a measure for the sharpness of
the unit:
\begin{eqnarray}\label{sin}
\langle\sin^3\theta\rangle\ \ \ \left\{\begin{array}{ll}
\rightarrow 1 &
\textrm{infinitely sharp to the left}\\
=0 & \textrm{symmetric upon reflection about
$y$-axis}\\
\rightarrow -1 & \textrm{infinitely sharp to the right}.
\end{array}\right.
\end{eqnarray}
On the other hand the quantity $0\leq \sitgem\leq 1$ is proportional to the linear friction coefficient, cf.  Eq.
(\ref{gammageneral}). For a bar of zero width, it has the minimal value $\sitgem=0$ when
the bar is oriented horizontally, while it is maximal ($\sitgem=1$) for a vertical bar.

\begin{table}
\centering
\begin{tabular}{|c|c|c|}
\hline \textrm{Shape} & \textrm{Fig.}  & \textrm{Stationary
velocity $V_0$ (order $\ep$)}
 \\
 \hline
 & &  \\
\textrm{Triangula}   &  \ref{fig:models}(a)  & $\frac{\sqrt{2 \pi
k_B m}}{4M}(1-\sinul)
\frac{\rho_1\rho_2(T_1-T_2)(\sqrt{T_1}-\sqrt{T_2})}
{\left[\rho_1\sqrt{T_1}+\rho_2\sqrt{T_2}\right]^2}$
\\ & & \\
\textrm{Triangulita}   & \ref{fig:models}(b)   & $\frac{\sqrt{2
\pi k_B
m}}{4M}(1-\sitnul)\frac{2\rho_1\rho_2\sqrt{T_1}(T_1-T_2)}{\left[2\rho_1
\sqrt{T_1}+\rho_2 \sqrt{T_2}(1+\sinul)\right]^2}$
\\ & & \\
\textrm{Triangle - triangle ($\rhd\lhd$) } & \ref{fig:models}(c) &
$\frac{\sqrt{2 \pi k_B m}}{4M}(1-\sinul)
\frac{\rho_1\rho_2(T_1-T_2)(\sqrt{T_1}+\sqrt{T_2})}
{\left[\rho_1\sqrt{T_1}+\rho_2\sqrt{T_2}\right]^2}$
\\ & & \\
\textrm{Triangle - triangle ($\theta_1\neq\theta_2$)} &
\ref{fig:models}(d) & $\frac{\sqrt{2 \pi k_B m}}{4M}
\frac{\rho_1\rho_2 (1+\sin\theta_1)(1+\sin\theta_2)
(T_1-T_2)[\sqrt{T_1}(1-\sin\theta_2)-\sqrt{T_2}(1-\sin\theta_1)]}
{\left[\rho_1\sqrt{T_1}(1+\sin\theta_1)+\rho_2\sqrt{T_2}(1+\sin\theta_2)\right]^2}$
\\ & & \\
 \hline
\end{tabular}
\caption{Analytic result for the lowest order contribution to
$\Vgem$ for the different constructions depicted in Fig.
\ref{fig:models}.} \label{tab:models}
\end{table}

\begin{figure}
\begin{center}
\includegraphics[width=10cm]{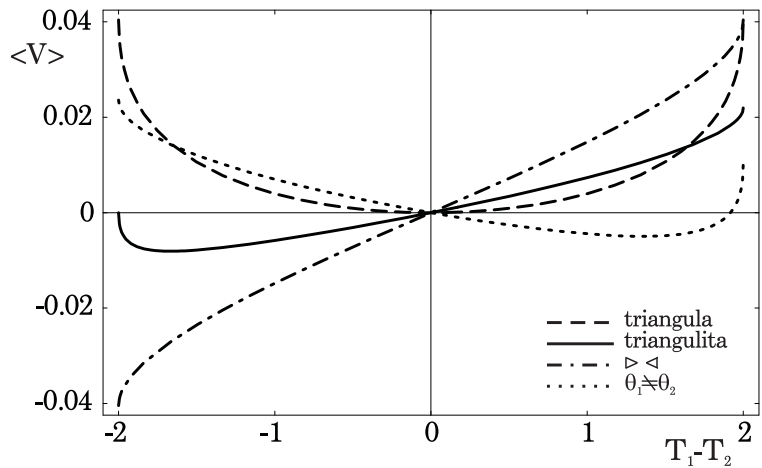}
\caption{Plot of the average speed (see Table \ref{tab:models})
for the four different models depicted in Fig. \ref{fig:models} as
a function of the applied temperature difference. The following
parameter values are used: $M/m=20$, $\rho_1=\rho_2=0.0022$,
$\theta_0=\pi/36$, $\theta_1=\pi/1000$, $\theta_2=\pi/3$ and
$T_1+T_2=2$ fixed.}\label{fig:fluxreversal}
\end{center}
\end{figure}

Based on Eq. (\ref{sgn}),  one can distinguish four different
classes of motors, depending on how the direction of  speed is
affected upon interchanging the temperatures $T_1$ and $T_2$. The
examples of Fig. \ref{fig:models} each represent one such class.
Fig. \ref{fig:fluxreversal} illustrates their behavior upon
switching $T_1$ and $T_2$.

\section{Systematic speed: special cases}

To gain further insight on the direction of the systematic speed, it is instructive
to calculate its explicit value for a number of special cases, see also Table
\ref{tab:models}.

\subsection{Identical units and nonlinear response}
For a construction with identical motor units, see for example
Triangula (Fig. \ref{fig:models}(a)) with two triangles pointing
in the same direction, Eq. (\ref{sgn}) reduces to
$\sgn(V_0)=\sgn(-\langle\sin^3\theta\rangle)$. One concludes that
the direction of Triangula's motion is independent of the
temperature gradient. This behavior originates from the
permutational symmetry of the identical motor units, implying that
the speed must be invariant under the interchange of $T_1,\rho_1$
with $T_2,\rho_2$. Note that the average velocity is pointing in
the direction of  the sharp angle of the motor units. It turns out
that this feature is general: for a motor with two identical
units, Eq. (\ref{eq:general}) simplifies to
\begin{eqnarray}
V_0=\frac{\sqrt{2 \pi m k_B }}{4M}\rho_1\rho_2
\frac{(T_1-T_2)(\sqrt{T_1}-\sqrt{T_2})}{\left[\rho_1\sqrt{T_1}+\rho_2\sqrt{T_2}\right]^2}
\frac{\sidgem}{\sitgem}.\label{IU}
\end{eqnarray}
We conclude that the motion is always in the same direction, the
``easy'' direction determined by the sign of $\sidgem$, cf. Eq.
(\ref{sin}).

We again stress that while the above direction of motion appears
to  be plausible, in agreement with our intuition, this specific
feature is exceptional and a result of  the permutation  symmetry
of the units. In fact, the latter symmetry has a dramatic effect
on the response of the motor to the temperature gradient: as is
clear from Eq. (\ref{IU}), the speed has a parabolic profile in
function of the applied temperature difference $T_1-T_2$. In
particular there is no linear response regime, with a speed
proportional to $T_1-T_2$! It is due to this unusual, symmetry
induced property that equilibrium is not a point of flux reversal.


\subsection{Identical units, with opposite orientation}

Another interesting class of motors are those with identical
units, but pointing in opposite directions, cf. Fig.
\ref{fig:models}(d). For these motors is
$F_1(\theta)=F_2(\pi-\theta)$ and consequently
\begin{eqnarray}
\singeme=\left\{\begin{array}{ll}\singemt & \textrm{ n even}\\
-\singemt & \textrm{ n odd}.
\end{array}\right.
\end{eqnarray}
One finds  from Eq. (\ref{eq:general}) that the average velocity
is given by
\begin{eqnarray}
V_0=\frac{\sqrt{2 \pi m k_B }}{4M}\rho_1\rho_2
\frac{(T_1-T_2)(\sqrt{T_1}+\sqrt{T_2})}{\left[\rho_1\sqrt{T_1}+\rho_2\sqrt{T_2}\right]^2}
\frac{\sidgeme}{\sitgeme},
\end{eqnarray}
We conclude  that the equilibrium state $T_1=T_2$ is the (only)
point of flux reversal. In the particular case of equal density
($\rho_1=\rho_2$), we have a symmetric Brownian motor, i.e. the
absolute value of its speed is invariant upon inverting the
temperature gradient, cf. \cite{gomez}.

\subsection{Different units}
When the shape of the two motor units is different, the situation
can be quite more involved. In particular, there may  be multiple
points of flux reversal in function of the applied temperature
difference. This reinforces the statement that the direction of
motion is difficult to predict. As an explicit example, we show in
Fig. (\ref{fig:fluxreversal}) that a construction with two
different triangles (Fig. \ref{fig:models}(d)) has a second
reversal point for $T_1 \gg T_2$.

\subsection{Multiple reservoirs}

\begin{figure}
\begin{center}
\includegraphics[width=6cm]{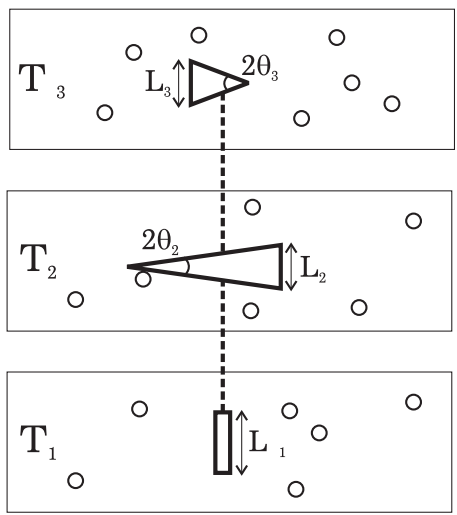} \caption{Construction of a thermal Brownian motor with $3$
reservoirs. The three motor units are rigidly linked and move as
one single entity.}\label{fig:3res}
\end{center}
\end{figure}

To illustrate the effect of multiple reservoirs on the amplitude and direction of the speed,
 we consider the case with three units,
namely two triangles (with different orientation and apex angle)
and a bar, see Fig. \ref{fig:3res}. The stationary velocity reads
\begin{eqnarray}
\fl V_0= \frac{\sqrt{2 \pi m k_B }}{4M}\frac{1}{\left[ 2\rho_1 L_1
\sqrt{T_1} + \rho_2 L_2 (1+\sin\theta_2)\sqrt{T_2} +\rho_3 L_3
(1+\sin\theta_3)\sqrt{T_3}
\right]^2}\nonumber\\
\fl \ \ \ \ \ \ \ \ \ \times \left\{
\rho_2(1-\sin^2\theta_2)\left[2\rho_1 L_1 \sqrt{T_1}
(T_2-T_1)+\rho_3 L_3 (1+\sin\theta_3)\sqrt{T_3}
(T_2-T_3)\right]  \right. \nonumber\\
\fl \ \ \ \ \ \ \ \ \ +\left.
 \rho_3(\sin^2\theta_3-1) \left[2\rho_1 L_1
\sqrt{T_1} (T_3-T_1)+\rho_2 L_2 (1+\sin\theta_2)\sqrt{T_2}
(T_3-T_2)\right]\right\}.
\end{eqnarray}
The speed is of the same order of magnitude as before.  The net
motion disappears at equilibrium ($T_1=T_2=T_3$), when the
asymmetry vanishes ($\theta_2=\theta_3=0$, both triangles are
flat), or when the effect of the two units cancels each other (for
example $\theta_2=\theta_3$, $T_2=T_3$ and $\rho_2=\rho_3$, i.e.,
both have the same opening angle in a reservoir at the same
temperature and with equal density).

One may wonder what is the effect of the addition of a reservoir
that is at equilibrium with a motor consisting of two units (at
lowest order in $\ep$). More precisely, consider  $T_1$  to be
equal to the effective temperature of reservoirs $2$ and $3$,
$T_1={(\gamma_2 T_2 + \gamma_3 T_3)}/{(\gamma_2+\gamma_3)}$.  We
find from Eq. (\ref{eq:general}):
\begin{eqnarray}
V_0&=& \frac{\sqrt{2 \pi m k_B }}{4M}\frac{\rho_2 S_2 (T_2-T_1)
\langle \sid\rangle_{_2}+\rho_3 S_3 (T_3-T_1) \langle
\sid\rangle_{_3}}{\sum_{i=1}^3 \rho_i S_i \sqrt{T_i} \langle
\sit\rangle_{_i} }.
\end{eqnarray}
We conclude that the presence of motor unit $1$ slows down the
systematic speed. Its single effect is the addition of an extra
friction term, cf. the denominator in the above formula.

\section{Efficiency}

As pointed out in the introduction, the efficiency of a thermal
motor, like the one discussed here, has to be compared with that
of a Carnot engine. The fact that we can extract work by adding a
load to the motor implies that there must be a concomitant
transfer of energy from the hot to the cold reservoir. A
fundamental difference with the Carnot construction is however
that the motor is at all times simultaneously in contact with both
the thermal reservoirs (we consider for simplicity the case of two
reservoirs). As a result, we expect that the object will acquire a
temperature intermediate between $T_1$ and $T_2$. This observation
is confirmed by the calculations mentioned above, where it is
found that, to lowest order in $\ep$, the velocity distribution of
the motor is Gaussian at an effective temperature $\Te$  given by
Eq. (\ref{teff}). At this  order of perturbation, one can easily
evaluate the corresponding heat flux between the reservoirs and
show that it obeys a Fourier law, see below. Furthermore, it has
been a matter of debate whether this irreversible entropy
producing process is reducing (dramatically) the efficiency of the
motor as a thermal engine. We will not discuss this matter
further, see \cite{parrondo2}, \cite{jarzynski}, \cite{sokolov}
and \cite{vandenbroeckprl} for more details . We rather focus on a
more practical definition of efficiency, namely the one considered
in finite time thermodynamics. Indeed, in practical applications,
Carnot efficiency is a less relevant quantity, since the central
issue is not work but power. Hence, we will evaluate the
efficiency of the Brownian motor at maximum power. To do so, we
have to repeat the above analysis but including an explicit
calculation of the heat flux and applying a (constant) load force
$F$ to the motor.

\subsection{Heat conduction}

To explicitly include the energy transfer between the heat
reservoirs, we add in the Boltzmann-Master equation, an additional
variable $E_1$ corresponding to the energy (change) in reservoir
$1$. Adding the direct acceleration produced by the external
force, we obtain the following equation for the probability
density $P(V,E_1,t)$ (prime is denoting precollisional values):
\begin{eqnarray}\label{boltzmannE}
\fl \partial_t P(V,E_1,t)&=&\int dV' W_1(V|V') P(V',E_1',t)
-\int dV' W_1(V'|V) P(V,E_1,t)\nonumber\\
& &+\int dV' W_2(V|V') P(V',E_1,t)
-\int dV' W_2(V'|V) P(V,E_1,t)\nonumber\\
& &-\frac{\partial}{\partial V}\left(\frac{F}{M}P(V,E_1,t)\right),
\end{eqnarray}
with $W(V'|V)=\sum_i W_i(V'|V)$ the transition probability per
unit time for the motor to go from speed $V'$ to $V$, given by Eq.
(\ref{eq:W}). Conservation of energy in reservoir $1$ implies
\begin{eqnarray}
\frac{1}{2}M V^2 + E_1&=& \frac{1}{2}M V'^2 + E_1^\prime.
\end{eqnarray}
For the transition probability $W_i(V;r)\equiv W_i(V'|V)$ with
jump amplitude $r=V'-V$, the conservation of energy permits us to
rewrite the above equation as:
\begin{eqnarray}
\fl \partial_t P(V,E_1,t)&=&\int dr W_1(V-r;r)
P(V-r,E_1+\frac{1}{2}Mr(2V-r),t) \nonumber\\& &
-\int dr W_1(V;r) P(V,E_1,t)\nonumber\\
& &+\int dr W_2(V-r;r) P(V-r,E_1,t)
-\int dr W_2(V;r) P(V,E_1,t)\nonumber\\
& &-\frac{\partial}{\partial V}\left(\frac{F}{M}P(V,E_1,t)\right).
\end{eqnarray}
Equivalently, we have:
\begin{eqnarray}
\fl
\partial_t P(V,E_1,t)
 =\int dr
\left[e^{-r\frac{\partial}{\partial V}}e^{\frac{M
r}{2}(2V+r)\frac{\partial}{\partial E_1}}-1\right]W_1(V;r)
P(V,E_1,t)\nonumber\\
+\left[e^{-r\frac{\partial}{\partial
V}}-1\right]W_2(V;r)P(V,E_1,t)-\frac{\partial}{\partial
V}\left(\frac{F}{M}P(V,E_1,t)\right).
\end{eqnarray}
Since $\langle E_1 \rangle=\int dV\int dE_1 E_1 P(V,E_1,t)$, the
energy exchange between the reservoirs obeys the following
equation:
\begin{eqnarray}
\fl
\partial_t\langle E_1 \rangle= \int dr\int dV\int dE_1 E_1
\left\{\left[\sum_{n=0}^{\infty}
\frac{(-r)^n}{n!}\frac{\partial^n}{\partial
V^n}\sum_{m=0}^{\infty}\frac{(\frac{M
r}{2}(2V+r))^m}{m!}\frac{\partial^m}{\partial
E_1^m}-1\right]W_1(V;r)\right.\nonumber\\
+ \left.\left[\sum_{n=0}^{\infty}
\frac{(-r)^n}{n!}\frac{\partial^n}{\partial V^n}-1\right]W_2(V;r)
-\frac{\partial}{\partial V}\frac{F}{M} \right\} P(V,E_1,t) .
\end{eqnarray}
The only non-vanishing contribution comes from the term $n=0,
m=1$, whence
\begin{eqnarray}
\partial_t\langle E_1 \rangle &=& -\langle \int dr
\frac{M r }{2} (2V+r) W_1(V;r) \rangle.
\end{eqnarray}
Using the jump moments $a_{n,i}(V)$ defined by Eq. (\ref{defAn}),
we can write for the heat flux from reservoir 1 to reservoir 2:
\begin{eqnarray}
\dot{Q}_{1\rightarrow 2}  = - \partial_t \langle E_1
\rangle&=&M\langle V a_{1,1}(V)\rangle +\frac{M}{2}\langle
a_{2,1}(V) \rangle.
\end{eqnarray}
Again, the functions $a_{1,1}$ and $a_{2,1}$, cf. Eqs.
(\ref{a1})-(\ref{a2}), can be expanded in terms of
$\ep=\sqrt{m/M}$ to find the corresponding expansion for the heat
flux in the limit of large mass $M$. To lowest order, one finds
that the energy flux obeys a Fourier law
\begin{eqnarray}
\dot{Q}_{1\rightarrow 2}=\kappa(T_1-T_2) +
O\left(\frac{1}{M^2}\right),
\end{eqnarray}
with
\begin{eqnarray}
\kappa=\frac{k_B}{M}\frac{\gamma_1\gamma_2}{\gamma_1+\gamma_2}.
\end{eqnarray}

\subsection{Power}

The time evolution for the average speed  obtained from the
Boltzmann equation (\ref{boltzmannE}),
\begin{eqnarray}\label{moment1F}
\partial_t \langle V\rangle&=& \langle a_1(V)\rangle +
\frac{F}{M},
\end{eqnarray}
differs from Eq. (\ref{seteq}) by the acceleration resulting from
the addition of the force $F$. To lowest order in $\ep$, the
steady state solution of Eq. (\ref{seteq2}) for the second moment
is unaffected by $F$ (i.e., $M\langle V^2\rangle=k_B\Te$) and we
find that
\begin{eqnarray}
\langle a_1(V)\rangle = -\frac{\gamma}{M}(\langle V\rangle -V_0).
\end{eqnarray}
Here is $\gamma$ the friction coefficient, cf. Eq.
(\ref{gammageneral}), and $V_0$ the average steady state velocity
in absence of an external force, see Eq. (\ref{eq:general}). The
steady-state solution of Eq. (\ref{moment1F}) thus reads:
\begin{eqnarray}\label{VF}
\langle V\rangle&=&V_0+\frac{F}{\gamma}.
\end{eqnarray}
Note that $F_{stop}=-\gamma V_0$ is the stopping force (at this
order of perturbation). Obviously, in order to extract work from
the Brownian motor we need to apply a force smaller than
$F_{stop}$. In particular one has that $F\sim 1/M$.


One defines the efficiency of the motor as the delivered work per
unit time over the heat exchange per unit time (for simplicity, we
assume from here on that $T_1>T_2$)
\begin{eqnarray}
\eta&=&\frac{P}{\dot{Q}_{1\rightarrow 2}}=\frac{-F\langle
V\rangle}{\dot{Q}_{1\rightarrow 2}}.
\end{eqnarray}
Since to lowest order of the perturbation in $\ep$ the heat flux
is not influenced by an external constant force $F$, we can
maximize the efficiency by maximizing the power $-F\langle
V\rangle$. One easily verifies that the latter is maximal when we
apply a force equal to half the stopping force,  $F=-\gamma
V_0/2$.
\begin{table}
\centering
\begin{tabular}{|c|c|c|}
\hline \textrm{Shape} & \textrm{Fig.}  & \textrm{Efficiency $\eta$
at maximum power}
 \\
 \hline
 & &  \\
\textrm{Triangula}   &  \ref{fig:models}(a) & $\frac{\pi}{32}
\frac{m}{M} (1-\sin{\theta_0})^2
\frac{(\sqrt{T_1}-\sqrt{T_2})^3}{\sqrt{T_1
T_2}(\sqrt{T_1}+\sqrt{T_2})}$
\\ & & \\
\textrm{Triangulita}   &  \ref{fig:models}(b)   & $\frac{\pi}{16}
\frac{m}{M}
\frac{(1-\sin^2{\theta_0})^2}{1+\sin{\theta_0}}\sqrt{\frac{T_1}{T_2}}
\frac{T_1-T_2}{(2\sqrt{T_1}+(1+\sin\theta_0)\sqrt{T_2})^2}$
\\ & & \\
\textrm{Triangle - triangle ($\rhd\lhd$)} & \ref{fig:models}(c) &
$\frac{\pi}{32} \frac{m}{M} (1-\sin{\theta_0})^2
\frac{T_1-T_2}{\sqrt{T_1 T_2}}$
\\ & & \\
\textrm{Triangle - triangle ($\theta_1\neq\theta_2$)} &
\ref{fig:models}(d) & $\frac{\pi}{32} \frac{m}{M}
(1+\sin\theta_1)(1+\sin\theta_2)\frac{T_1-T_2}{\sqrt{T_1
T_2}}\left[\frac{
\sqrt{T_1}(1-\sin\theta_2)-\sqrt{T_2}(1-\sin\theta_1)}
{\sqrt{T_1}(1+\sin\theta_1)+\sqrt{T_2}(1+\sin\theta_2)}\right]^2$
\\ & & \\
 \hline
\end{tabular}
\caption{Analytic result for the lowest order contribution to the
efficiency $\eta$ at maximum power, in the situation of equal
densities in the reservoirs ($\rho_1=\rho_2$,
$T_1>T_2$).}\label{tab:efficiency}
\end{table}
The resulting efficiency of the thermal engine at maximum power
is then given by:
\begin{eqnarray}\label{eff}
\eta=\frac{M}{4
k_B}\frac{(\gamma_1+\gamma_2)^2}{\gamma_1\gamma_2}\frac{V_0^2}{T_1-T_2}.
\end{eqnarray}
One  can immediately deduce from Eqs. (\ref{gammageneral}) and
(\ref{eq:general}), that the efficiency $\eta$ is very low, due to
a prefactor $m/M$: for small $m/M$ the heat conductivity is the
dominant factor and both the rectified motion and resulting work
are a second order effect. This observation is of course
restricted to our discussion of the small $\ep$ limit. In fact
much higher efficiencies can be in principle be achieved when one
moves away from this limit \cite{vandenbroeck}.

Since the expression for the efficiency of an arbitrary motor, Eq.
(\ref{eff}), is rather complex, we have reproduced the efficiency
for the models of Fig. \ref{fig:models} in Table
\ref{tab:efficiency}. In particular, $\eta$ simplifies
significantly for identical conditions in both reservoirs (equal
densities $\rho_1=\rho_2$ and equal objects
$F_1(\theta)=F_2(\theta)$ and $S_1=S_2$):
\begin{eqnarray}
\eta&=&\frac{\pi}{32}\frac{m}{M}\left(\frac{\langle
\sin^3\theta\rangle}{\langle \sin^2\theta\rangle}\right)^2
\frac{(\sqrt{T_1}-\sqrt{T_2})^3}{\sqrt{T_1T_2}(\sqrt{T_1}+\sqrt{T_2})}.
\end{eqnarray}


\end{document}